\documentclass[aps,reprint,superscriptaddress,nofootinbib,preprintnumbers,longbibliography]{revtex4-2} 
\usepackage{url}
\usepackage{upgreek,amssymb,multirow}
\usepackage{graphicx,color}
\usepackage{mathtools}
\usepackage{float}
\usepackage{mciteplus}
\usepackage{blindtext}
 
\usepackage{amsmath}
\usepackage{extarrows}
\usepackage{braket}
\usepackage{tikz}
\usepackage{tikz-cd}

\usepackage{amsthm}
\theoremstyle{plain}
\newtheorem{prop}{Proposition}
\newtheorem*{prop*}{Proposition}
\newtheorem{cor}[prop]{Corollary}
\newtheorem*{cor*}{Corollary}
\newtheorem{claim}[prop]{Claim}
\usepackage[hyperindex,breaklinks]{hyperref}
\usepackage[title,titletoc]{appendix}
\definecolor{Myblue}{rgb}{0,0,0.9}
\hypersetup{
    linktocpage=true,
    colorlinks=true,
    citecolor=blue,
    linkcolor=blue,
    urlcolor=blue,
}

\begin{document}
\title{Self-\texorpdfstring{\(G\)}{G}-ality in 1+1 dimensions}
\author{Takamasa Ando}

\affiliation{Center for Gravitational Physics and Quantum Information, Yukawa Institute for Theoretical Physics, Kyoto University, Kyoto 606-8502, Japan} 
\begin{abstract}
    We explore topological manipulations in one spatial dimension, which are defined for a system with a global symmetry and map the system to another one with a dual symmetry. In particular, we discuss fusion category symmetries enhanced by the invariance of the actions of topological manipulations, i.e.~self-\(G\)-alities for topological manipulations. Based on the self-\(G\)-ality conditions, we provide LSM-type constraints on the ground states of many-body Hamiltonians. We clarify the relationship between different enhanced symmetries and how they are further enhanced when they meet. We explore concrete lattice models for such self-\(G\)-alities and identify how the self-\(G\)-ality structures match the IR critical theories.
\end{abstract}
    
    
\maketitle
    
\tableofcontents

\section{Introduction}\label{sec:intro}
Symmetry plays a crucial role in physics. Depending on the global symmetries of the system, we can consider the classification of quantum phases of matter, namely ground states of many-body Hamiltonians. According to the Landau paradigm, some quantum phases are classified by spontaneous symmetry breaking (SSB). In recent decades, the theory of quantum phases that are not captured by SSB has been developed. Symmetry-protected topological (SPT) phases \cite{Gu:2009dr, Pollmann:2009mhk, Pollmann:2009ryx, Chen:2010zpc, Schuch:2010, Chen:2011pg, Levin:2012yb} are examples of such quantum phases. SPT phase is a class of quantum phases that has a unique gapped ground state. In general, wide classes of gapped quantum phases (not necessarily SPTs) are described by topological quantum field theories (TQFTs).

Recently the understanding of global symmetries has been intensively developing. The global symmetry of a theory is the data of topological operators, i.e., operators whose expectation values are invariant under continuous deformations of the operators, in the theory. A large amount of work on such \emph{generalized} symmetries has been done \cite{Kapustin:2014gua, Gaiotto:2014kfa, Bhardwaj:2017xup, Chang:2018iay, Thorngren:2019iar, Choi:2021kmx, Kaidi:2021xfk, Bhardwaj:2022yxj, Delcamp:2023kew}. Notably, symmetry operators do not necessarily have inverse elements, and such operators are said to be non-invertible. In general, algebraic data of topological operators are captured by (higher) categories. In this paper, we refer to categories that describe the data of global symmetries as \textit{symmetry categories}. In one spatial dimension, finite generalized symmetries are described by fusion categories \cite{Bhardwaj:2017xup, Chang:2018iay, Thorngren:2019iar}. Usual group symmetries are also included in categorical symmetry formalism. For instance, finite \(G\) group symmetry is described by \(\mathrm{Vec}_{G}^{\omega}\), the category of \(G\)-graded finite dimensional \(\mathbb{C}\) vector spaces whose associator is specified by a three-cocycle \(\omega\). The system with a non-vanishing \(\omega\) is said to have an 't Hooft anomaly.
Classification of (1+1)-dimensional quantum phases with fusion category symmetries has also been done. It is shown that unitary \((1+1)d\) TQFTs with fusion category \(\mathcal{C}\) symmetries are classified by module categories over \(\mathcal{C}\) \cite{Thorngren:2019iar, Komargodski:2020mxz}. See also \cite{Huang:2021zvu, Inamura:2021szw}.

For a given global symmetry, one can define \textit{topological manipulations}, which map quantum phases to another one. Specifically, elements of topological manipulations consist of gauging (gaugeable) algebra and stacking SPTs \cite{Gaiotto:2020iye, Kaidi:2022cpf, Aksoy:2022hua, Bhardwaj:2022maz, Bhardwaj:2022kot}. One example of topological manipulations is the Kramers-Wannier (KW) transformation, which is defined for systems with \(\mathbb{Z}_2\) global symmetries in 1+1 dimensions. This transformation is understood as gauging the \(\mathbb{Z}_2\) global symmetries. It exchanges two gapped phases with a \(\mathbb{Z}_2\) symmetry, one is the \(\mathbb{Z}_2\) spontaneously broken phase and the other is without SSB phase. Explicit implementation of the KW transformation on the closed chain has been developed in the literature \cite{Li:2023ani, Chen:2023qst, Seiberg:2024gek}. While gauging algebras are often realized by non-local transformation, stacking \((1+1)d\) bosonic SPTs are implemented by finite-depth local unitary (FDLU) quantum circuits \cite{Chen:2010gda}. One of the features of topological manipulations is that they are in general \emph{non-invertible} transformations, i.e., there is no inverse element. 

Mathematical structures of topological manipulations are well described by Symmetry Topological Field Theory (SymTFT) \cite{Thorngren:2019iar, Ji:2019jhk, Kong:2020cie, Gaiotto:2020iye, Aasen:2020jwb, Freed:2022qnc}.\footnote{In condensed matter literature, SymTFT is also called a symmetry topological order or topological holography.} SymTFT is a \((d+1)\)-dimensional TQFT associated to \(d\)-dimensional systems with a symmetry category \(\mathcal{C}\). Starting from the SymTFT, one can construct the original \(d\)-dimensional system by the so-called sandwich construction. In SymTFT picture, topological manipulations are described by invertible surface operators.\footnote{Not all topological manipulations can be described by invertible surfaces of SymTFT. In this paper, we only consider topological manipulations that are captured by invertible surfaces in SymTFT.} We review this in Sec.~\ref{sec:SymTFT}. Based on this circumstance, in this paper we adopt composition rule of topological manipulations to be that of the corresponding surface operators in SymTFTs. Then a set of topological manipulations forms a finite group. In particular, for any topological manipulation \(f\), there is an integer \(n\) such that \(n\)-th power of \(f\) is equal to the identity operation. Such manipulation \(f\) is referred to as an \textit{\(n\)-ality operation}. When \(n=2\) and \(n=3\), we refer to each as a \textit{duality operation} and a \textit{triality operation}, respectively. The KW transformation is an example of duality operations.

A system can be invariant under the action of topological manipulations that form a group \(G\). The condition for invariance under such manipulations is called a \textit{self-\(G\)-ality} condition. When a system is self-\(G\)-al, the symmetry category is \emph{enhanced} to a larger symmetry category in the sense that the original category is a subcategory of the enhanced symmetry category. In the example of KW transformation, the symmetry group \(\mathbb{Z}_2\) is enhanced to the Tambara-Yamagami category \cite{TAMBARA1998692} \(\mathrm{TY}(\mathbb{Z}_2,+)\). From the point of view of the field-theory formalism, one expects that self-\(G\)-ality conditions impose strong constraints on the ground states of the system. While constraints imposed by the self-duality for the KW transformation have been explored in \cite{Levin:2019ifu,Seiberg:2024gek}, constraints for general symmetries on the lattice have not been fully developed.

Understanding how self-\(G\)-ality structures occur in physical systems is also a crucial problem. In physics situations, self-dualities appear or emerge at phase transition points between distinct quantum phases. 
To illustrate, consider a topological manipulation that exchanges two gapped phases. Take a one-parameter family of Hamiltonians for the two phases, in which each Hamiltonian in the family is mapped to another one. Then suppose that the phase transition between the two phases occurs at a point in the family. Since the two gapped phases cannot match the self-duality condition, the IR theory at the critical point should exhibit the enhanced symmetry category. Therefore, identifying enhanced symmetry categories can provide a helpful approach to studying critical systems. Even though, study of concrete lattice models for critical phases specified by self-\(G\)-alities is not enough.%
\footnote{See, e.g., \cite{Bhardwaj:2024wlr, Bhardwaj:2024kvy} for related work on generalized symmetries in critical systems.}

In this paper, we explore various self-\(G\)-alities and their dynamical consequences. In particular, we discuss self-\(G\)-alities and consequences for systems with \(\mathbb{Z}_2\times\mathbb{Z}_2\) symmetries. 
We also discuss the web of enhanced symmetry categories and the further enhancement of symmetries, which are often found in multicritical theories.
To see self-\(G\)-ality structures more explicitly, we study critical lattice models that symmetry enhancement exactly appears at the UV scale and figure out how the self-\(G\)-ality structures match the IR critical theories.
Though the lattice model with a \(\mathbb{Z}_2 \times \mathbb{Z}_2\) symmetry has already been studied in previous work, we explicitly identify the global symmetries of the underlying IR theories in terms of enhanced symmetries and clarify the relation between the critical IR theories. We also study how these symmetries are matched at the multicritical point where the self-duality lines meet. The analysis uses only lattice transformations that can be understood in terms of the SymTFT or the topological manipulation framework, and so can be applied to other lattice models. See Sec.~\ref{sec:Z2Z2_model} for details of the comparison with previous studies.

\subsection*{Overview of the paper}
In Sec.~\ref{sec:SymTFT}, we briefly review SymTFTs and explain how the KW transformation is described by a surface operator in the SymTFT. 

In Sec.~\ref{sec:Z2Z2_manip} and \ref{sec:Z2Z2_enhance}, we introduce and review topological manipulations in \((1+1)\)-dimensional \(\mathbb{Z}_2\times\mathbb{Z}_2\) symmetric systems and their symmetry enhancement studied in the literature. In \ref{sec:no-go}, we provide various LSM-type arguments for the enhanced symmetry categories by looking at the behavior of order parameters under the topological manipulations. 
In Sec.~\ref{sec:dual_dual}, we clarify the web of dualities between enhanced symmetries and discuss its physical mechanism.
Based on such dualities between self-dualities, in Sec.~\ref{sec:multicritical} we study how symmetries are further enhanced when some self-dual theories intersect. 

In Sec.~\ref{sec:Z2Z2_model}, we analyze a concrete lattice model for the enhanced symmetry categories introduced in Sec.~\ref{sec:Z2Z2}. We analytically solve the model and find that self-duality lines are described by the compact boson conformal field theory (CFT), and discuss how self-\(G\)-ality structures are matched in the IR CFT. 

In Sec.~\ref{sec:ZnZn}, we discuss duality operations in \((1+1)\)-dimensional \(\mathbb{Z}_n\times\mathbb{Z}_n\) symmetric systems. The self-duality for the operations can be understood as the enhanced symmetry of the critical phase between two distinct SPT phases. 

Appendix \ref{sec:Z2Z2_QFT} is about the review of the formulation by field theories. We discuss symmetry enhancement by stacking general \((1+1)d\) SPTs in Appendix \ref{sec:enhance_by_SPT}. In Appendix \ref{sec:cptbsn}, we detail the analysis of the compact boson for the lattice model in Sec.~\ref{sec:Z2Z2_model}.

\section{Topological manipulation and Symmetry TFT}\label{sec:SymTFT}
In this section, we overview how topological manipulations are implemented in SymTFTs. In particular, we see how the KW transformation is realized in the corresponding SymTFT. As explained in the introduction, SymTFT is a framework for studying the kinematic/algebraic data of symmetry categories by using TQFTs in one higher dimension. Specifically, SymTFT for \((1+1)d\) systems with a \(\mathcal{C}\) fusion category symmetry is realized by the Turaev-Viro-Barrett-Westbury (TVBW) construction \cite{TURAEV1992865, Barrett:1993ab}. \((2+1)d\) theory obtained by this construction is a TQFT, whose anyon data is captured by the Modular tensor category (MTC) \(\mathcal{Z}(\mathcal{C})\) of the Drinfeld center of \(\mathcal{C}\).

Let us briefly review the sandwich construction for obtaining the \(\mathcal{C}\) symmetric theory from SymTFTs. First, consider a \((d+1)\)-dimensional spacetime manifold \(M_{d}\times [0,1]\) with boundary conditions on \(M_{d}\times\{0\}\) and \(M_{d}\times\{1\}\). We choose one boundary to be a \textit{physical} boundary where the original \(d\)-dimensional theory lives, and the other to be \textit{topological} boundary carrying the symmetry category \(\mathcal{C}\). Since the bulk SymTFT is topological one can shrink the bulk and obtain the original theory with a \(\mathcal{C}\) symmetry defined on \(M_{d}\). When one take the physical boundary to be topological, the \(d\)-dimensional theory obtained by this sandwich construction is a TQFT with \(\mathcal{C}\) symmetry. In general, a topological boundary condition of a \((2+1)d\) TQFT with a MTC \(\mathcal{B}\) is realized by condensing a Lagrangian algebra of \(\mathcal{B}\) \cite{Fuchs:2012dt}. In the SymTFT picture, some topological manipulations are interpreted as actions of codimension-one operators in the bulk TQFT on physical boundaries. 

Let us give an example. SymTFT for non-anomalous \(\mathbb{Z}_2\) symmetry in \((1+1)d\) is the untwisted \(\mathbb{Z}_2\) Dijkgraaf-Witten theory \(D(\mathbb{Z}_2)\). This TQFT is also known as the toric code phase. Anyons in \(D(\mathbb{Z}_2)\) are generated by following four simple anyons:
\begin{equation}
    \{1,e,m,f\mid e^2=m^2=f^2=1\},
\end{equation}
where \(e\) and \(m\) are bosons, and \(f\) is a fermion. Namely, the topological spin of \(e,m,f\) is \(1,1,-1\), respectively. In \(D(\mathbb{Z}_2)\), there are two topological boundaries. One is the Dirichlet boundary condition, which is realized by condensing two anyons \(1\) and \(e\). The other is the Neumann boundary condition, whose condensed anyons are \(1\) and \(m\). As the topological boundary of the SymTFT, we take the Dirichlet boundary. Since we have two topological boundary conditions, there are two TQFTs with a non-anomalous \(\mathbb{Z}_2\) symmetry. When one takes the Dirichlet/Neumann boundary as the physical boundary, one obtains the \(\mathbb{Z}_2\) SSB/trivial phase. The KW transformation is a topological manipulation that exchanges the two phases. In SymTFT picture, this manipulation is realized by a surface operator that acts the four anyons as 
\begin{equation}
    1\mapsto 1,\quad e\mapsto m,\quad m\mapsto e,\quad f\mapsto f.
\end{equation}
Such a surface operator \(S_{em}\) are realized by condensing the fermion \(f\). Moreover, one can verify that the surface operator is invertible, in particular satisfies \(S_{em}^2=1\), see e.g.~\cite{Roumpedakis:2022aik, Kaidi:2022cpf} for the derivation. From this discussion one can find that the KW transformation is a duality operation.

\section{\texorpdfstring{\(\mathbb{Z}_2\times\mathbb{Z}_2\)}{Z2 x Z2} symmetry}\label{sec:Z2Z2}
In this section, we discuss self-\(G\)-alities under topological manipulations for \(\mathbb{Z}_2\times\mathbb{Z}_2\) symmetries. Though we take a specific representation of \(\mathbb{Z}_2\times\mathbb{Z}_2\) group, the discussion in this section can be applied to any system if the dimension of the local Hilbert space is large enough to define KW like transformations.

Consider a one-dimensional spin-1/2 chain of size \(2L\). We define generators of a \(\mathbb{Z}_2^A\times\mathbb{Z}_2^B\) as
\begin{equation}\label{Z2Z2_sym}
    U_A\coloneqq \prod_{j:\text{odd}}\sigma_{j}^x,\quad  U_B\coloneqq \prod_{j:\text{even}}\sigma_{j}^x.
\end{equation}

\subsection{Gapped phases and topological manipulations}\label{sec:Z2Z2_manip}
There are six (stable) gapped phases with a \(\mathbb{Z}_2\times\mathbb{Z}_2\) symmetry. Five of these are distinguished by spontaneous symmetry-breaking patterns, and the remaining two are distinct SPT phases. To characterize these six gapped phases, we use order parameters. Each SSB is diagnosed by the following two-point correlation functions:
\begin{align}\label{order_SSB}
    \begin{split}
        W_{\text{SSB}}^A(j,j+r)&=\sigma_{j}^z\sigma_{j+r}^z, \quad (j:\text{odd})\\ 
        W_{\text{SSB}}^B(j,j+r)&=\sigma_{j}^z\sigma_{j+r}^z, \quad (j:\text{even}). 
    \end{split}
\end{align}
When expectation values of these order parameters with respect to the ground state of the system behave as \(\langle W^{A/B}_{\text{SSB}}(j,j+r)\rangle\neq 0\) in the limit of \(r\nearrow\infty\), the corresponding \(\mathbb{Z}_2^{A/B}\) is spontaneously broken. Note that SSB to the diagonal \(\mathbb{Z}_2\) subgroup is diagnosed by \(W_{\text{SSB}}^A(j,j+r)W_{\text{SSB}}^B(j+1,j+r+1)\).

To distinguish two SPT phases, we can use string order parameters. The trivial phase is characterized by the two string operators of the form
\begin{equation}\label{order_triv}
    W_0^A(j,j+r)=\prod_{\substack{0\leq i\leq r,\\j+i\,:\,\text{odd}}}\sigma_{j+i}^x, \quad  
        W_0^B(j,j+r)=\prod_{\substack{0\leq i\leq r,\\j+i\,:\,\text{even}}}\sigma_{i}^x,
\end{equation}
and the nontrivial phase is characterized by\footnote{The string order parameter that diagnoses the nontrivial \(\mathbb{Z}_2\times\mathbb{Z}_2\) SPT phase in a spin-one chain was introduced in \cite{denNijs:1989ntw}.}
\begin{align}\label{order_SPT}
    \begin{split}
        W_{\text{SPT}}^A(j,j+r)&=\sigma_{j-1}^z\left(\prod_{\substack{0\leq i\leq r,\\j+i\,:\,\text{odd}}}\sigma_{j+i}^x\right)\sigma_{j+r+1}^z,\\ 
        W_{\text{SPT}}^B(j,j+r)&=\sigma_{j-1}^z\left(\prod_{\substack{0\leq i\leq r,\\j+i\,:\,\text{even}}}\sigma_{j+i}^x\right)\sigma_{j+r+1}^z.
    \end{split}
\end{align}
When a ground state belongs to the trivial/(nontrivial) SPT phase, both the expectation values \(W_{\text{0/SPT}}^A\) and \(W_{\text{0/SPT}}^B\) for the ground state take non-vanishing values under \(r\nearrow \infty\).
The six gapped phases considered here are stable phases under generic perturbations preserving the \(\mathbb{Z}_2\times\mathbb{Z}_2\) symmetry. Mathematically, they correspond to indecomposable module categories over \(\mathrm{Vec}_{\mathbb{Z}_2\times\mathbb{Z}_2}\), the category of \(\mathbb{Z}_2\times\mathbb{Z}_2\) graded vector spaces. More general gapped phases are given by direct sum of these.

In the rest of this section, we focus on three gapped phases, the trivial phase (Trivial), the nontrivial SPT phase (SPT), and the full \(\mathbb{Z}_2^A\times\mathbb{Z}_2^B\) SSB phase (SSB). The web of dualities of the three gapped phases are explored in \cite{Thorngren:2019iar, Li:2023ani}, see Fig.~\ref{fig:web_Z2Z2}.

Let us introduce two topological manipulations \(S\) and \(T\): gauging the \(\mathbb{Z}_2^A\times\mathbb{Z}_2^B\) symmetry and stacking an SPT. The topological manipulation \(S\) is implemented on the lattice as 
\begin{align}\label{Z2Z2_S}
    \begin{split}
        \sigma_{2j-1}^x\mapsto \sigma_{2j-2}^z\sigma_{2j}^z, \quad \sigma_{2j-1}^z\sigma_{2j+1}^z\mapsto \sigma_{2j}^x,\\
    \sigma_{2j}^x\mapsto \sigma_{2j-1}^z\sigma_{2j+1}^z, \quad \sigma_{2j-2}^z\sigma_{2j}^z\mapsto \sigma_{2j-1}^x.
    \end{split}
\end{align}
see \cite{Li:2023ani, Seifnashri:2024dsd} for the details of this transformation.

The topological manipulation \(T\) by stacking a nontrivial SPT phase is implemented as
\begin{equation}\label{Z2Z2_T}
    \sigma_{j}^x\mapsto \sigma_{j-1}^z\sigma_{j}^x\sigma_{j+1}^z.
\end{equation}
This transformation is for example realized by
\begin{equation}\label{T_Z4dual}
    V_T\coloneqq \prod_{j}e^{\frac{\pi i}{4}\left(1+(-1)^j\sigma_{j}^z\sigma_{j+1}^z\right)}.
\end{equation}
Note that the unitary operator that realizes \eqref{Z2Z2_T} is not unique.

SymTFT of \(\mathbb{Z}_2\times\mathbb{Z}_2\) symmetries is the untwisted \(\mathbb{Z}_2\times\mathbb{Z}_2\) Dijkgraaf-Witten theory \(D(\mathbb{Z}_2\times\mathbb{Z}_2)\). Anyons in \(D(\mathbb{Z}_2\times\mathbb{Z}_2)\) are labeled as \(\{e_1^{a_1}e_2^{a_2}m_1^{b_1}m_2^{b_2}\mid a_1,a_2,b_1,b_2=0,1\}\). Three topological manipulations discussed in this section are realized as \(\mathbb{Z}_2\) invertible surface operators in \(D(\mathbb{Z}_2\times\mathbb{Z}_2)\). To find the surface operator corresponding to the manipulation \(S\), we first consider the \(e_1\leftrightarrow m_1\) and \(e_2\leftrightarrow m_2\) exchange operator \(\widetilde{S}\), which is realized by condensing two fermions \(e_1m_1,e_2m_2\) independently. Specifically, \(\widetilde{S}\) acts on each anyon as 
\begin{equation}
    e_1^{a_1}e_2^{a_2}m_1^{b_1}m_2^{b_2}\mapsto e_1^{b_1}e_2^{b_2}m_1^{a_1}m_2^{a_2}.
\end{equation}
The combination of \(\widetilde{S}\) and the label \(1\leftrightarrow 2\) exchange symmetry gives the bulk \(\mathbb{Z}_2\) symmetry corresponding to the manipulation \(S\).

The surface operator that realizes the manipulation \(T\) is constructed by higher-gauging \cite{Roumpedakis:2022aik} of \(m_1\) and \(m_2\) anyons with the nontrivial discrete torsion.
The explicit action of the surface operator for the \(T\) operation on anyons are given by 
\begin{equation}
    m_1\mapsto e_2m_1, \quad m_2\mapsto e_1m_2, \quad e_1\mapsto e_2, \quad e_2\mapsto e_1.
\end{equation}
The others are determined by the composition. In \(D(\mathbb{Z}_2\times\mathbb{Z}_2)\), there are many other surface operators and they are explored in \cite{Fuchs:2014ema, Gaiotto:2020iye}.

By the above SymTFT picture, one finds that the manipulations \(S\) and \(T\) are both duality operations, i.e., their square is equivalent to the identity operation. Combining \(S\) and \(T\), we obtain another duality operation, which is \(STS\). This duality operation has been known as the Kennedy-Tasaki (KT) transformation \cite{kennedy1992hidden, Kennedy:1992ifl,Oshikawa_1992, Else:2013gsf, Duivenvoorden:2013tfa}, which was originally considered in spin-1 chains.

Though \(S\) and \(T\) are duality operations, their composition \(ST\) and \(TS\) are not duality operations but \textit{triality} operations. One can easily see the triality structure by the definitions \eqref{Z2Z2_S} and \eqref{Z2Z2_T}. A set \(\{1,S,T,STS,ST,TS\}\) forms a non-Abelian group \(S_3\) of the symmetric group on a set of three elements. Note that the triality of \(ST\) requires \(STS=TST\) as a duality operation.

\subsection{Enhanced symmetry category}\label{sec:Z2Z2_enhance}
In this subsection, we discuss symmetry categories enhanced by topological manipulations defined in the previous subsection. First, we study the enhancement by \(S,T,\) and \(STS\).
\begin{itemize}
    \item \(S\) induces the exchange Trivial \(\leftrightarrow\) SSB, and the SPT phase is self-dual. When the system is self-dual for \(S\) operation, the global symmetry is enhanced to \(\mathrm{TY}(\mathbb{Z}_2^A\times\mathbb{Z}_2^B,\chi,+)\) of Tambara-Yamagami (TY) category with an off-diagonal bicharacter \(\chi\).\footnote{For \((a_i,b_i)_{i=1,2}\in \{0,1\}\), \(\chi\) is defined as \(
        \chi((a_1,b_1),(a_2,b_2))\coloneqq (-1)^{a_1b_2+b_1a_2}.\)} This fusion category is equivalent to \(\mathrm{Rep}(D_8)\) of the representation category of the order-eight Dihedral group \(D_8\).
    \item \(STS\) induces the exchange SPT \(\leftrightarrow\) SSB, and trivial phase is self-dual. The enhanced symmetry category with a self-duality \(STS\) is equivalent to  \(\mathrm{TY}(\mathbb{Z}_2^A\times\mathbb{Z}_2^B,\chi,+)\). 
    \item  \(T\) induces the exchange Trivial \(\leftrightarrow\) SPT, and the SSB phase is self-dual. The enhanced symmetry category is the pointed category \(\mathrm{Vec}_{\mathbb{Z}_2^A\times\mathbb{Z}_2^B\times\mathbb{Z}_2^C}^\omega\) of \(\mathbb{Z}_2^A\times\mathbb{Z}_2^B\times\mathbb{Z}_2^C\) graded vector spaces with a three-cocycle \(\omega=(-1)^{\int ABC}\). Physically, the system with this symmetry category exhibits a mixed anomaly.
\end{itemize}
The self-duality and the enhanced symmetry category for \(S\) operation were well-discussed in the literature \cite{Thorngren:2019iar, Li:2023ani, Seifnashri:2024dsd}. The symmetry category \(\mathrm{Rep}(D_8)\) admits three fiber functors, i.e.~there are three distinct SPT phases with a \(\mathrm{Rep}(D_8)\) symmetry. Lattice models that realize these phases are discussed in \cite{Seifnashri:2024dsd}. The second enhanced symmetry category is studied in \cite{Li:2023ani}. In particular, the authors computed the fusion rules and verified the rule is TY fusion category of \(\mathbb{Z}_2^A\times\mathbb{Z}_2^B\). 
The third enhanced symmetry category is discussed in \cite{Tantivasadakarn:2021wdv, Seifnashri:2024dsd}. The \(\mathbb{Z}_2^C\) symmetry in the third is generated by \(T\), and an SPT entangler on the lattice is given by \eqref{T_Z4dual}. From the expression \eqref{T_Z4dual}, we explicitly verify that the \(\mathbb{Z}_2^A\times\mathbb{Z}_2^B\times\mathbb{Z}_2^C\) exhibits a mixed anomaly with \(\omega=(-1)^{\int ABC}\). To see this we first note that the symmetry does not have any anomalies with type-one, i.e.~pure \(\mathbb{Z}_2\) anomaly because the two generators \eqref{Z2Z2_sym} are on-site, and \eqref{T_Z4dual} is only consists of \(\sigma_{j}^z\). To see the cocycle \(\omega=(-1)^{\int ABC}\), we consider the diagonal \(\mathbb{Z}_2\) subgroup of \(\mathbb{Z}_2^A\times\mathbb{Z}_2^B\times\mathbb{Z}_2^C\). Then we obtain \(\prod_{j}\sigma_{j}^x\prod_{j}\exp\left(\frac{\pi i}{4}(1-(-1)^{j}\sigma_{j}^z\sigma_{j+1}^z)\right)\) for the generator, and this is local-unitary equivalent to the known \(\mathbb{Z}_2\) anomalous symmetry action \cite{Levin:2012yb}. Moreover, we can also see that there is no type-two anomaly by using the Else-Nayak procedure \cite{Else:2014vma}, which is reviewed in App.~\ref{sec:EN}. We give a more formal way to understand the mixed anomaly in Sec.~\ref{sec:dual_dual}.
\paragraph*{Remark.}
    Symmetry categories enhanced by stacking SPTs do not depend on how one chooses the FDLU \(V\). In general, we specify symmetry categories enhanced by stacking general SPTs without assuming any concrete expression of FDLUs in App.~\ref{sec:enhance_by_SPT}.
\paragraph*{Remark.}
    When considering the self-duality for gauging the \(\mathbb{Z}_2^A\times\mathbb{Z}_2^B\) symmetry, we have another choice of a topological manipulation \(S^\prime\). It is given by 
    \begin{align}\label{Z2Z2_S2}
        \begin{split}
            \sigma_{2j-1}^x\mapsto \sigma_{2j-1}^z\sigma_{2j+1}^z, \quad \sigma_{2j-1}^z\sigma_{2j+1}^z\mapsto \sigma_{2j+1}^x,\\
        \sigma_{2j}^x\mapsto \sigma_{2j}^z\sigma_{2j+2}^z, \quad \sigma_{2j}^z\sigma_{2j+2}^z\mapsto \sigma_{2j+2}^x.
        \end{split}
    \end{align}
    Note that \(S\) and \(S^\prime\) are equivalent up to relabeling the two \(\mathbb{Z}_2\) global symmetries at the level of topological manipulations. Nevertheless, the difference is crucial when considering self-dualities. The extended symmetry category with a self-duality for \(S^\prime\) is \(\mathrm{TY}(\mathbb{Z}_2^A\times\mathbb{Z}_2^B,\chi^\prime,+)\). Here the bicharacter \(\chi^\prime\) is a diagonal one.\footnote{\(\chi^\prime\) is defined as \(\chi((a_1,b_1),(a_2,b_2))\coloneqq (-1)^{a_1a_2+b_1b_2}.\)} This TY category is known as \(\mathrm{Rep}(H_8)\)\footnote{\(H_8\) is a non-group like Hopf algebra of dimension eight known as the Kac-Paljutkin algebra\cite{Thorngren:2019iar}.} 
    and it is a subcategory of \(\mathrm{TY}(\mathbb{Z}_2,+)\boxtimes\mathrm{TY}(\mathbb{Z}_2,+)\), see further \cite{Thorngren:2019iar}. \(\mathrm{Rep}(H_8)\) symmetry in the compact boson is studied in \cite{Choi:2023vgk}.

\paragraph*{Other enhanced symmetry category.}
As explained in the previous subsection, \(ST\) and \(TS\) are triality operations, and the group generated by \(S\) and \(T\) is a non-Ableian group \(S_3\). We do not specify fusion categories enhanced by these manipulations. Instead, we discuss how such enhanced categories appear in a concrete theory in the next section.

\subsection{No-go argument}\label{sec:no-go}
If a system satisfies a self-\(G\)-ality condition, some gapped phases are forbidden due to the condition. In condensed matter literature, such no-go arguments, derived from the properties of the UV scale, are called LSM-type constraints. The name comes from the LSM theorem, which has been extensively studied \cite{Lieb:1961fr, Oshikawa:2000zwq, Hastings:2003zx, Hastings_2005, Watanabe:filling, Ogata:2020hry, Yao:2020xcm, Aksoy:2021uxb, Aksoy:2023hve}. LSM-type constraint for the KW transformation is explored in \cite{Levin:2019ifu, Seiberg:2024gek}.
Here, we provide LSM-type constraints for systems with self-duality conditions for manipulations in the \(\mathbb{Z}_2\times\mathbb{Z}_2\) symmetry \eqref{Z2Z2_sym}. We note that the no-go statements discussed here are almost trivial at the level of TQFTs. They immediately follow from the structure of topological manipulations in Fig.~\ref{fig:web_Z2Z2}. Nevertheless, the criterion on the lattice is nontrivial, so we discuss no-go statements for lattice models here. We also remark that the discussion here is not mathematically rigorous, whereas we expect it to be physically reasonable.

We have seen that the self-duality for \(T\) operation induces the enhanced anomalous symmetry \(\mathrm{Vec}_{\mathbb{Z}_2^A\times\mathbb{Z}_2^B\times\mathbb{Z}_2^C}^{\omega}\). Famously, 't Hooft anomalies forbid the existence of unique gapped ground states, and such a no-go argument for this mixed anomaly was shown in \cite{Tantivasadakarn:2021wdv,  Li:2022nwa}, see \cite{Garre-Rubio:2022uum, Seifnashri:2023dpa, Rubio:2024tsw} for anomalies of general groups. In the following, we give an explanation for this and see other no-go statements based on a different approach: using order parameters.

We summarize the no-go statements in Table \ref{tb:no-go_Z2Z2}.
\begin{table*}[tbp]
    \centering
    \begin{tabular}{|c|c|c|c|} \hline
        Self-duality & Symmetry category & Forbidden phases & Stable gapped phases\\ \hline
        \(S\) & \(\mathrm{TY}(\mathbb{Z}_2^A\times\mathbb{Z}_2^B,\chi,+)\) & Trivial, SSB & SPT, \(\text{Trivial}\oplus\text{SSB}\), \(\ldots\)\\ 
        \(STS\) & \(\mathrm{TY}(\mathbb{Z}_2^A\times\mathbb{Z}_2^B,\chi,+)\) & SPT, SSB & Trivial, \(\text{SPT}\oplus\text{SSB}\), \(\ldots\) \\ 
        \(T\) & \(\mathrm{Vec}_{\mathbb{Z}_2^A\times\mathbb{Z}_2^B\times\mathbb{Z}_2^C}^{\omega}\) & Trivial, SPT &  SSB, \(\text{Trivial}\oplus\text{SPT}\), \(\ldots\) \\ \hline
    \end{tabular}
    \caption{Forbidden and stable gapped phases under self-dualities. Here, \(\oplus\) denotes a macroscopic superposition (direct sum) of many-body ground states. We do not list all allowed gapped phases. The complete list can be obtained easily by following the discussion in the main text.}
    \label{tb:no-go_Z2Z2}
\end{table*}
We use three order parameters introduced in \eqref{order_SSB}, \eqref{order_triv}, \eqref{order_SPT}. Recall that three gapped phases are characterized by the expectation values of the order parameters. We only show the first line in the table, i.e., allowed gapped phases under the invariance of \(S\).
The other statements follow from similar discussions. First, let us see how the order parameters are mapped under the topological manipulations \(S,STS,\) and \(T\). By the definitions of topological manipulations, one sees that 
\begin{equation}
    S\colon \begin{pmatrix}
        W_0^{A/B}(j,j+r)\\W_{\text{SPT}}^{A/B}(j,j+r)\\W_{\text{SSB}}^{A/B}(j,j+r)
    \end{pmatrix}
    \mapsto
    \begin{pmatrix}
        W_{\text{SSB}}^{B/A}(j-1,j+r+1)\\W_{\text{SPT}}^{A/B}(j,j+r)\\W_0^{B/A}(j+1,j+r-1)
    \end{pmatrix},
\end{equation}
\begin{equation}
    STS\colon \begin{pmatrix}
        W_0^{A/B}(j,j+r)\\W_{\text{SPT}}^{A/B}(j,j+r)\\W_{\text{SSB}}^{A/B}(j,j+r)
    \end{pmatrix}
    \mapsto
    \begin{pmatrix}
        W_{0}^{A/B}(j,j+r)\\ W_{\text{SSB}}^{A/B}(j-1,j+r+1)\\W_{\text{SPT}}^{A/B}(j+1,j+r-1)
    \end{pmatrix},
\end{equation}
\begin{equation}
    T\colon \begin{pmatrix}
        W_0^{A/B}(j,j+r)\\W_{\text{SPT}}^{A/B}(j,j+r)\\W_{\text{SSB}}^{A/B}(j,j+r)
    \end{pmatrix}
    \mapsto
    \begin{pmatrix}
        W_{\text{SPT}}^{A/B}(j,j+r)\\ W_0^{A/B}(j,j+r)\\W_{\text{SSB}}^{A/B}(j,j+r)
    \end{pmatrix}.
\end{equation}
Consider (possibly degenerate) ground states of an \(S\)-invariant Hamiltonian, and take one of them. If this state belongs to the trivial phase, then it has nonzero expectation values of \(W_{0}^{A/B}\), while the expectation values of the other order parameters vanish. Since the \(S\) operation maps \(W_{0}^{A/B}\) to \(W_{\mathrm{SSB}}^{A/B}\), the corresponding ground state obtained by acting with \(S\) on the original trivial state must have nonzero expectation values of \(W_{\mathrm{SSB}}^{A/B}\). Thus, the ground-state sector must include an SSB state. Conversely, any SSB state is mapped to a trivial state, which means that any trivial state and SSB state must appear as a pair. The minimal choice is simply a macroscopic superposition of a trivial state and an SSB state. On the other hand, the nontrivial SPT state can be a unique gapped ground state of an \(S\)-symmetric Hamiltonian because the order parameters \(W_{\mathrm{SPT}}^{B/A}\) are mapped back to themselves under the \(S\) operation. General allowed gapped states are given by macroscopic superpositions of these gapped states. We thus obtain the first line of Table~\ref{tb:no-go_Z2Z2}.

\subsection{Dualities between self-dualities}\label{sec:dual_dual}
As discussed above, we have three self-dualities with symmetry categories \(\mathcal{C}_{S}\coloneqq \mathrm{TY}(\mathbb{Z}_2^A\times\mathbb{Z}_2^B,\chi,+), \mathcal{C}_{STS}\coloneqq \mathrm{TY}(\mathbb{Z}_2^A\times\mathbb{Z}_2^B,\chi,+), \mathcal{C}_{T}\coloneqq \mathrm{Vec}_{\mathbb{Z}_2^A\times\mathbb{Z}_2^B\times\mathbb{Z}_2^C}^\omega\). We further have topological manipulations between these theories. The relation between \(\mathrm{Rep}(D_8)(\cong \mathrm{TY}(\mathbb{Z}_2\times\mathbb{Z}_2,\chi,+))\) and \(\mathrm{Vec}_{\mathbb{Z}_2^3}^{\omega}\) or \(\mathcal{C}_{S}\) and \(\mathcal{C}_T\) are well-known by the previous studies \cite{Tachikawa:2017gyf, Seifnashri:2024dsd}. In this subsection, we clarify the full structure of the duality web and discuss a general picture. 

Schematically, three enhanced symmetry categories are related as
\begin{equation}\label{dual_dual_Z2Z2}
    \mathcal{X}(\mathcal{C}_{S})\quad 
    \begin{tikzpicture}
        \draw[->,line width=.5pt] (0,0) -- (1.2,0);
        \draw[->,line width=.5pt] (1.2,0) -- (0,0);
        \node[scale=.9] at (1.2/2,0.3) {\(T\)};
    \end{tikzpicture}
    \quad \mathcal{X}({\mathcal{C}_{STS}})\quad 
    \begin{tikzpicture}
        \draw[->,line width=.5pt] (0,0) -- (1.2,0);
        \draw[->,line width=.5pt] (1.2,0) -- (0,0);
        \node[scale=.9] at (1.2/2,0.3) {\(S\)};
    \end{tikzpicture}
    \quad \mathcal{X}(\mathcal{C}_{T}),
\end{equation}
where \(\mathcal{X}(\mathcal{C}_\bullet )\) denotes the theory with \(\mathcal{C}_{\bullet}\) symmetry. The \(SL(2,\mathbb{Z}_2)\) structure in Fig.~\ref{fig:web_Z2Z2} plays a crucial role to \eqref{dual_dual_Z2Z2}. For example, take a theory \(\mathcal{X}(\mathcal{C}_{S})\) then we see that 
\begin{equation}
    T(ST(\mathcal{X}(\mathcal{C}_{S})))=STS(\mathcal{X}(\mathcal{C}_{S}))=ST(\mathcal{X}(\mathcal{C}_{S})),
\end{equation}
which means the theory \(ST(\mathcal{X}(\mathcal{C}_{S}))\) has \(\mathcal{C}_{T}\) symmetry. Here, we used the relation \(STS=TST\) in the first equality.\footnote{The author thanks Yunqin Zheng for telling this calculation to the author.}

Another way to see the relation \eqref{dual_dual_Z2Z2} is to start from the theory with a \(\mathrm{Vec}_{D_8}\) symmetry. The group \(D_8\) fits into the following central extension of groups:
\begin{equation}
    1\to \mathbb{Z}_2^C \to D_8 \to \mathbb{Z}_2^A\times\mathbb{Z}_2^B \to 1,
\end{equation}
where the second cohomology that specifies the extension is \(e\left((a_1,b_1),(a_2,b_2)\right)=(-1)^{a_1b_2}\). After gauging the \(\mathbb{Z}_2\) subgroup, we have a theory with a symmetry category \(\mathrm{Vec}_{\mathbb{Z}_2^A\times\mathbb{Z}_2^B\times\mathbb{Z}_2^C}^{\omega}=\mathcal{C}_{T}\) \cite{Tachikawa:2017gyf}. Since the original theory with a \(\mathrm{Vec}_{D_8}\) symmetry is non-anomalous, we can further gauge \(\mathbb{Z}_2^A\times\mathbb{Z}_2^B\), then we obtain the theory with a symmetry category \(\mathrm{Rep}(D_8)\cong \mathrm{TY}(\mathbb{Z}_2^A\times\mathbb{Z}_2^B,\chi,+)=\mathcal{C}_{STS}\). Since this \(\mathbb{Z}_2^A\times\mathbb{Z}_2^B\) gauging corresponds to the \(S\) operation (up to stacking an SPT), we see the web \eqref{dual_dual_Z2Z2}.

Can we consider topological manipulations between self-\(G\)-alities in more general settings? Suppose that \(\mathcal{X}_1,\mathcal{X}_2\) be invariant under the action of group of topological manipulation \(F_1,F_2\), respectively. The sufficient condition for \(\mathcal{X}_1\) and \(\mathcal{X}_2\) to be related via a topological manipulation is that \(F_1\) and \(F_2\) are conjugate with each other, i.e., there is a topological manipulation \(g\) such that \(F_1=g\circ F_2\circ g^{-1}\). Clearly, two groups \(F_1\) and \(F_2\) are necessarily isomorphic. Mathematically, the manipulation \(g\) gives a Moria equivalence between two symmetry categories enhanced by \(F_1\cong F_2\). We discuss this point from the point of view of SymTFT in the following.

When a physical boundary is invariant under the action of a set of surface operators (the group is denoted by \(G\)) in the SymTFT, one can consider the bulk theory with arbitrary insertions of the surface operators. Then the surface operators are transparent in the bulk theory, and the bulk theory becomes gauged theory by the \(G\) zero-form symmetry. In the last of Sec.~\ref{sec:dual_dual}, we discussed the conditions for the existence of topological manipulations that map two enhanced symmetry categories and explained mathematically such manipulations give the Morita equivalences between the enhanced categories. Now the physical interpretation of the equivalence is clear, namely topological manipulations between self-\(G\)-alities give automorphisms of SymTFTs. In particular, if two sets of surface operators are in the same conjugacy class of the group of invertible surface operators, which is known to be the Brauer-Picard group \cite{Etingof:2009yvg} of the symmetry category, in the SymTFT, the gauged TQFTs are the same.

\subsection{Multicriticality}\label{sec:multicritical}
As explained in the Introduction, self-\(G\)-ality structures naturally appear at critical points between gapped phases with an original symmetry. The global symmetries of such critical points are enhanced by self-\(G\)-alities. We can go one step further. As discussed in Sec.~\ref{sec:dual_dual}, there may also be topological manipulations between self-\(G\)-alities, and these manipulations can define maps between self-\(G\)-ality theories. Then the enhanced symmetries are further enhanced by invariance under such additional manipulations. Physically, this situation can often be found when two distinct critical lines, with different self-\(G\)-ality structures, intersect with each other, namely at multicritical points.%
\footnote{Symmetry structures do not enforce criticality (gaplessness) in general. Here we focus only on systems that are invariant under the action of some topological manipulations, although we use the terminology ``critical'' in this paragraph.}
Mathematically, such manipulations between two (or more) self-\(G\)-alities can be specified by Morita equivalences between enhanced symmetry categories. Indeed, the three enhanced symmetry categories in the \(\mathbb{Z}_2 \times \mathbb{Z}_2\) example are related by topological manipulations as in \eqref{dual_dual_Z2Z2}, where the two manipulations \(S\) and \(T\) give Morita equivalences among the three symmetry categories.

To see how enhanced symmetries are further enhanced at multicritical points, let us consider two sets of manipulations, \(\{F_{1,i}\}_i\) and \(\{F_{2,j}\}_j\), where \(i\) and \(j\) label elements of topological manipulations with respect to the original symmetry category \(\mathcal{C}\). The once-enhanced theories are invariant under the actions of \(F_{1,i}\) and \(F_{2,j}\), respectively. Now consider a further enhancement by topological manipulations \(\{G_k\}_k\); namely, the system is invariant under the actions of the \(G_k\). Since the \(G_k\) relate the once-enhanced systems, the resulting twice-enhanced system is invariant under \(\{F_{1,i}, F_{2,j}, G_k\}_{i,j,k}\), and this invariance determines the enhanced symmetry category.

Let us give a simple example of such multicritical theories. Consider two decoupled transverse-field Ising (TFI) Hamiltonian of the form 
\begin{align}
    \begin{split}
        H[J_A,J_B]=-\sum_{j}&\left(\sigma_{2j-1}^x+\sigma_{2j}^x+J_A\,\sigma_{2j-1}^z\sigma_{2j+1}^z\right.\\
        &\left.+J_B\,\sigma_{2j}^z\sigma_{2j+2}^z\right).
    \end{split}
\end{align}
This model is self-dual under gauging the \(\mathbb{Z}_2^{A/B}\) symmetry when \(J_{A/B}=1\). In both cases, the enhanced symmetry category is \(\mathrm{TY}(\mathbb{Z}_2,+)\boxtimes\mathrm{Vec}_{\mathbb{Z}_2}\). When one writes down the phase diagram with parameters \(J_A\) and \(J_B\), one obtains two critical transition lines, which are described by the Ising CFT with the central charge \(1/2\). These two lines intersect at the point \(J_A=J_B\), which is described by the two decoupled Ising CFT, and this point gives an example of multicriticalities. Here, the symmetry is enhanced from \(\mathrm{TY}(\mathbb{Z}_2,+)\boxtimes\mathrm{Vec}_{\mathbb{Z}_2}\) to \(\mathrm{TY}(\mathbb{Z}_2,+)\boxtimes\mathrm{TY}(\mathbb{Z}_2,+)\). We see another example of a codimension-two transition in the next section.

\section{Lattice models for self-\texorpdfstring{\(G\)}{G}-alities with \texorpdfstring{\(\mathbb{Z}_2\times\mathbb{Z}_2\)}{Z2 x Z2} symmetry}\label{sec:Z2Z2_model}
In this section, we explore a family of lattice models with the \(\mathbb{Z}_2\times\mathbb{Z}_2\) symmetry \eqref{Z2Z2_sym}. In particular, we discuss how the self-\(G\)-ality structures are matched with the lattice model.

The Hamiltonian we consider is the following:
\begin{equation}\label{Z2Z2_ham}
    H[J_0,J_1,J_S]=J_0H_0+J_1H_{\text{SPT}}+J_SH_{\text{SSB}},
\end{equation}
where \(J_0,J_1,J_S\) are real parameters and 
\begin{align}
    H_0&=-\sum_{j=1}^{L}\left(\sigma_{2j-1}^x+\sigma_{2j}^x\right),\\
    H_{\text{SPT}}&=-\sum_{j=1}^{L}\left(\sigma_{2j-2}^z\sigma_{2j-1}^x\sigma_{2j}^z+\sigma_{2j-1}^z\sigma_{2j}^x\sigma_{2j+1}^z\right),\\
    H_{\text{SSB}}&=-\sum_{j=1}^{L}\left(\sigma_{2j-1}^z\sigma_{2j+1}^z+\sigma_{2j}^z\sigma_{2j+2}^z\right).
\end{align}
When two parameters are the same, the Hamiltonian satisfies the self-duality condition under duality operation \(S/T/STS\). When \(J_0=J_1=J_S\), it is self-dual for both \(S\) and \(T\), and hence this point is self-\(S_3\)-ality point.

\begin{figure}[tbp]
    \centering
    \includegraphics[scale=1]{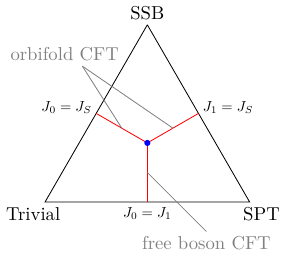}
    \caption{The phase diagram of the Hamiltonian \eqref{Z2Z2_ham}. Three red lines are self-duality lines, and the blue point in the center is self-\(S_3\)-ality point and it is described by the KT CFT.}
    \label{fig:Z2Z2_phase}
\end{figure}

In Fig.~\ref{fig:Z2Z2_phase}, we summarize the phase diagram of the family \eqref{Z2Z2_ham} in the range of \(-1<J_0,J_1,J_S\leq 1\). 

\paragraph*{Comparison with previous studies.}
Before going on to the analysis of the Hamiltonian \eqref{Z2Z2_ham}, let us comment on a comparison with previous studies. Part of the analysis of the structure of the duality web and the model \eqref{Z2Z2_ham} in this subsection are overlapped with the previous studies \cite{Verresen:2019igf} and \cite{Moradi:2022lqp}. In \cite{Verresen:2019igf}, the authors analyzed different but IR-equivalent models to ours and arrived at the same phase diagrams as Fig.~\ref{fig:Z2Z2_phase}. In \cite{Moradi:2022lqp}, the authors explored the same model as \eqref{Z2Z2_ham} and arrived at the same conclusion. However, the way to solve the model is different from ours. 
What we present in this paper is the specification of the IR CFTs on the three critical self-duality lines and the identification of how the enhanced symmetries are matched in these CFTs. We also expect the concrete way of solving the model to be helpful, because we use only lattice transformations, such as the KW transformation, which can be well understood as topological manipulations.

\subsection{Self-duality lines}
Let us examine the ground states of the self-duality lines. First, we see \(H[J_0=J_1, J_S]\). The other two lines can be mapped from this line by topological manipulations, and are discussed later. When \(J_0=J_1=1\), \(H[J_0, J_1, J_S]\) becomes 
\begin{equation}\label{dualXXZ}
    H[1,1,J_S]=-\sum_{j=1}^{2L}\left(\sigma_{j}^x+\sigma_{j-1}^z\sigma_{j}^x\sigma_{j+1}^z+J_S\sigma_{j}^z\sigma_{j+2}^z\right).
\end{equation}
Here, we assume that \(L\in 2\mathbb{Z}\) for simplicity. This is on-site unitary equivalent to\footnote{Since on-site unitary transformations do not change any entangle structures of lattice models, such transformations do not affect our analysis.}
\begin{equation}
    H^\prime[1,1,J_S]=-\sum_{j=1}^{2L}\left(\sigma_{j}^x-\sigma_{j-1}^z\sigma_{j}^x\sigma_{j+1}^z-J_S\sigma_{j}^z\sigma_{j+2}^z\right).
\end{equation}
Consider the KW transformation of the form
\begin{equation}\label{KW_XXZ}
    \sigma_{j}^z\sigma_{j+1}^z\mapsto Z_{j},\quad \sigma_{j}^x\mapsto X_{j-1}X_{j},
\end{equation}
which corresponds to gauge the diagonal \(\mathbb{Z}_2\) subgroup of \(\mathbb{Z}_2^A\times\mathbb{Z}_2^B\) symmetry. This transformation is equivalent to the one studied in \cite{PhysRevB.46.3486}. Here, \(X_{j}\) and \(Z_{j}\) are usual Pauli matrices. After the transformation, the Hamiltonian becomes 
\begin{equation}
    H^g=-\sum_{j=1}^{2L}\left(X_{j}X_{j+1}+Y_{j}Y_{j+1}-J_S\,Z_{j}Z_{j+1}\right),
\end{equation}
which is equivalent to the following XXZ chain model:
\begin{equation}\label{XXZ_chain}
    H_{\text{XXZ}}[J_S]=\sum_{j=1}^{2L}\left(X_{j}X_{j+1}+Y_{j}Y_{j+1}+J_S\,Z_{j}Z_{j+1}\right).
\end{equation}
The dual \(\mathbb{Z}_2\) symmetry of the gauged model is generated by
\begin{equation}\label{XXZ_Z2sym}
    U^g=\prod_{j}Z_{j}.
\end{equation}
Upon the \(\mathbb{Z}_2\) gauging, the unitary operator \(V_T\) \eqref{Z2Z2_T} for the \(T\) operation is also changed. It becomes
\begin{equation}
    V_T^\prime\mapsto \prod_{j}e^{\frac{\pi i}{4}\left(1-Z_{j}\right)},
\end{equation}
which now generates a \(\mathbb{Z}_4\) symmetry and its \(\mathbb{Z}_2\) subgroup is generated by \eqref{XXZ_Z2sym}. This nontrivial symmetry extension reflects the structure of a mixed anomaly between \((\mathbb{Z}_2^A\times\mathbb{Z}_2^B)_{\text{diag.}}\) symmetry and the \(\mathbb{Z}_2\) symmetry generated by \(V_T\) \cite{Tachikawa:2017gyf}. The IR behavior of the XXZ chain \eqref{XXZ_chain} has been well-studied in the literature. By the exact solution by the Bethe ansatz, \eqref{XXZ_chain} is described by the compact boson CFT for \(-1<J_S\leq 1\), and the radius \(R^\prime\) of the corresponding compact boson is given by \cite{Luther:1975wr, PhysRevLett.45.1358}\footnote{Here, we chose the normalization of the radius so that \(R=\sqrt{2}\) is the \(SU(2)_1\) CFT.}
\begin{equation}
    (R^\prime)^2=\dfrac{2}{\left(1-\dfrac{1}{\pi}\mathrm{Arccos}(J_S)\right)}.
\end{equation}
The \(\mathbb{Z}_2\) symmetry \eqref{XXZ_Z2sym} corresponds to the \(\mathbb{Z}_2^S\) shift symmetry of the compact boson. When one gauges \(\mathbb{Z}_2^S\) the model comes back to \eqref{dualXXZ}, and the theory is also a compact boson with a radius
\begin{equation}\label{cpt_radius_Z2Z2}
    R^2=\dfrac{1}{2\left(1-\dfrac{1}{\pi}\mathrm{Arccos}(J_S)\right)}.
\end{equation}
By T-duality, the theory with a radius \eqref{cpt_radius_Z2Z2} is equivalent to the theory with a radius
\begin{equation}\label{cpt_radius_Tdual}
    R^2=8\left(1-\frac{1}{\pi}\mathrm{Arccos}(J_S)\right).
\end{equation}
When \(J_S=0\), \(R\) is \(2\), and the compact boson at this point is the free Dirac theory. This is consistent with the fact that we have the XX model when we set \(J_S=0\) in \eqref{XXZ_chain}. Precisely, when \(J_S=0\) both \(R\) and \(R^\prime\) are equal to \(2\), which means the free Dirac or the \(U(1)_4\) CFT is self-dual under \(\mathbb{Z}_2^S\) gauging. When \(J_S=1\), the radius is \(2\sqrt{2}\), and this point is described by the KT CFT \cite{Ginsparg:1987eb}. In the gauged picture \eqref{XXZ_chain}, \(J_S=1\) point is described by the \(SU(2)_1\) CFT.

In summary, when \(-1<J_S\leq 1\), \(H[1,1,J_S]\) is described by the compact boson CFT with a radius \eqref{cpt_radius_Z2Z2}. To identify the IR CFTs on the other two self-duality lines, we need to find how the global symmetry \eqref{Z2Z2_sym} acts on the compact boson CFT. We discuss it in the following subsection.

\subsection{Self-\texorpdfstring{\(G\)}{G}-ality structures}
The compact boson at a generic radius has a global symmetry \(\left(U(1)^S\times U(1)^W\right)\rtimes \mathbb{Z}_2^C\), where \(U(1)^{S/W}\) is the shift/winding symmetry and \(\mathbb{Z}_2^C\) is the charge conjugation symmetry.\footnote{Note that this \(\mathbb{Z}_2^C\) is different from the \(\mathbb{Z}_2\) symmetry generated by stacking a nontrivial SPT in the previous section.} By an explicit calculation in App.~\ref{sec:cptbsn}, we find that the global symmetry \eqref{Z2Z2_sym} corresponds to \(\mathbb{Z}_2^S\subset U(1)^S\) and the \(\mathbb{Z}_2\) diagonal subgroup of \(\mathbb{Z}_2^S\times\mathbb{Z}_2^C\). If one gauges the \(\mathbb{Z}_2^A\times\mathbb{Z}_2^B\) symmetry, the compact boson is mapped to the orbifold branch. This is consistent with the fact that the Hamiltonian \eqref{Z2Z2_ham} becomes the two decoupled critical TFI, which is described by the two decoupled Ising CFT, when \(J_0=J_S,J_1=0\). The diagonal subgroup of the two \(\mathbb{Z}_2\) symmetries is \(\mathbb{Z}_2^S\), or it is \(\mathbb{Z}_2^W\) as the T-dual model. Then the KW transformation \eqref{KW_XXZ} can be interested as gauging the \(\mathbb{Z}_2^W\) symmetry of the compact boson at the radius \eqref{cpt_radius_Tdual}.
Since stacking an SPT does not change the dynamical properties of IR CFTs, we obtain the phase diagram in Fig.~\ref{fig:Z2Z2_phase}. The discussion here tells us that the orbifold branch of the compact boson at a generic radius has a \(\mathrm{TY}(\mathbb{Z}_2\times\mathbb{Z}_2,\chi,+)\cong\mathrm{Rep}(D_8)\) symmetry. \(\mathrm{Rep}(D_8)\) symmetry in the orbifold branch is discussed in \cite{Thorngren:2021yso}.

Self-duality for stacking an SPT forces the partition function of the theory to vanish when the configuration of the SPT phase gives a nontrivial phase factor \((-1)^{\int_{M_2}AB}=-1\). Here, \(M_2\) is a spacetime manifold, and \(A\) and \(B\) is background \(\mathbb{Z}_2\) gauge fields for \(\mathbb{Z}_2^{A}\) and \(\mathbb{Z}_2^A\) symmetry, respectively. 

We see this property holds in the compact boson CFT because the twisted partition function by \(\mathbb{Z}_2^C\) symmetry and \((\mathbb{Z}_2^S\times\mathbb{Z}_2^C)_{\text{diag.}}\) symmetry gives the same projection for the Hilbert space. Using this vanishing property, we explicitly confirm that the KT CFT is invariant under the triality operation \(ST\), see App.~\ref{sec:cptbsn}.
Combining this self-triality with the other self-dualities, we find that the KT CFT realizes a self-\(S_3\)-ality.

\begin{figure}
    \begin{tikzpicture}
        \centering
        \pgfmathsetmacro{\sepy}{2}
        \node[] at (0,0) {\(H_{\text{XXZ}}[J]\): circle CFT};
        \node[] at (0,-\sepy) {\(H[1,1,J]\): circle CFT};
        \node[] at (0.3,-\sepy) {};
        \node[] at (0,-2*\sepy) {\(H[1,J,1]\): orbifold CFT};
        \node[] at (0,-3*\sepy) {\(H_{\text{AT}}[J]\): orbifold CFT};

        \draw[->] (0.2,-0.3) -- node[scale=.8,anchor=west]{gauge \(\mathbb{Z}_2^S\)} (0.2,-\sepy+0.3);
        \draw[<-] (-0.2,-0.3) -- node[scale=.8,anchor=east]{gauge \(\mathbb{Z}_2^W\)} (-0.2,-\sepy+0.3);

        \draw[->] (0.2,-\sepy-0.3) -- node[scale=.8,anchor=west]{\(TS\)} (0.2,-2*\sepy+0.3);
        \draw[<-] (-0.2,-\sepy-0.3) -- node[scale=.8,anchor=east]{\(ST\)} (-0.2,-2*\sepy+0.3);

        \draw[->] (0.2,-2*\sepy-0.3) -- node[scale=.8,anchor=west]{gauge \(\mathbb{Z}_2^A\)} (0.2,-3*\sepy+0.3);
        \draw[<-] (-0.2,-2*\sepy-0.3) -- node[scale=.8,anchor=east]{gauge \(\mathbb{Z}_2^{\hat{A}}\)} (-0.2,-3*\sepy+0.3);
    \end{tikzpicture}
    \caption{The relation between the XXZ model and the Ashkin-Teller model.}
    \label{fig:web_of_XXZ-AT}
\end{figure}
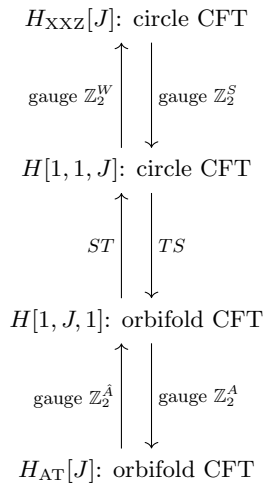

\paragraph*{Revisiting the Ashkin-Teller and XXZ.}
In \cite{Moradi:2022lqp}, the authors analyze \(H[1,J_1,1]\) in a different way. Starting from \(H[1,J_1,1]\), they gauge the \(\mathbb{Z}_2^A\) symmetry and obtain the model of the form 
\begin{align}\label{AT_ham}
    \begin{split}
        H_{\text{AT}}[J_1]=-\sum_{j=1}^{L}&\left(\sigma_{j}^x+\sigma_{j}^z\sigma_{j+2}^z+J_1\sigma_{2j-2}^z\sigma_{2j-1}^z\sigma_{2j}^z\sigma_{2j+1}^z\right.\\
        &\left.+J_1\sigma_{2j}^x\sigma_{2j+1}^x\right),
    \end{split}
\end{align}
which is known as the quantum Ashkin-Teller (AT) model. The relation between the XXZ model and the AT model is studied in \cite{KohmotoPRB81, Alcaraz:PhysRevLett.58.771, Shibata:2019cvd}. From the analysis we have done, we now clearly understand the relation of them, which is schematically in Fig.~\ref{fig:web_of_XXZ-AT}.

\section{\texorpdfstring{\(\mathbb{Z}_n\times\mathbb{Z}_n\)}{Zn x Zn} symmetry}\label{sec:ZnZn}
In \((1+1)d\), we have \(n\) distinct bosonic SPT phases with a \(\mathbb{Z}_n\times\mathbb{Z}_n\) symmetry. Lattice model for each SPT phase is for instance realized by the following Hamiltonian \cite{Chen:2011pg}:
\begin{equation}\label{eq:ZnZn_SPT}
    H_{k}=-\sum_{j}\left(Z_{2j}^kX_{2j+1}(Z_{2j+2}^\dagger)^k+(Z_{2j-1}^\dagger)^k X_{2j}Z_{2j+1}^k\right)+\mathrm{h.c.},
\end{equation}
where \(X_j,Z_j\) are \(n\)-level spin and satisfy \(X_{j}^n=Z_{j}^n=I_{n},Z_{j}X_{j}=e^{2\pi i/n}X_{j}Z_{j}\) (\(I_{n}\) is the \(n\times n\) identify matrix).

Similarly, we have a lattice model for the full \(\mathbb{Z}_n\times\mathbb{Z}_n\) SSB phase:
\begin{equation}\label{eq:ZnZn_SSB}
    H_{\text{SSB}}=-\sum_{j=1}^{L}\left(Z_{2j-1}Z_{2j+1}^\dagger+Z_{2j}Z_{2j+2}^\dagger\right)+\mathrm{h.c.}.
\end{equation}
The model \(H_{0,\text{SSB}}\coloneqq H_{0}+H_{\text{SSB}}\) describes the critical point of the two decoupled quantum clock models, and it is self-dual under the topological manipulation, which is a generalization of \(S\) in the previous sections.

\subsection{SPT transition}\label{sec:SPT_transition}
Consider a Hamiltonian \(H_{k,k^\prime}\coloneqq H_{k}+H_{k^\prime}\). For \(n=2,3,4\) and \(k=0\), this model is explored in \cite{Tsui:2017ryj}. When \(k-k^\prime\) is coprime to \(n\), we construct a (invertible) topological manipulation that maps \(H_{k,k^\prime}\) to \(H_{0,\text{SSB}}\) in App.~\ref{sec:ZnZn_QFT}. Since topological manipulations do not change the gaplessness in general, we find that \(H_{k,k^\prime}\) realizes the model for the critical phase between two SPT phases. In particular, the central charge of the IR CFT is the same as \(H_{0,\text{SSB}}\). To construct the manipulation, we prove the following statement:
\begin{claim}\label{claim}
    Consider two SPT phases with a \(\mathbb{Z}_n\times\mathbb{Z}_n\) symmetry of levels \(k\) and \(k^\prime\). There is a duality operation between these two SPT phases if \(k-k^\prime\) is coprime to \(n\).
\end{claim}
Physically, critical phases described by the Hamiltonian \(H_{k,k^\prime}\) are characterized by the symmetry category enhanced by the duality operation in the claim. It would be interesting to see how the structure of the enhanced symmetry category is encoded in the corresponding IR CFT or TQFT and we leave it for future research.

\section{Outlook}
In this work, we explored the space of topological manipulations in 1+1 dimensions, in particular with \(\mathbb{Z}_2\times\mathbb{Z}_2\) symmetries. There are several future directions:
\begin{itemize}
    \item Mathematically, an enhancement symmetry category by a group \(G\) topological manipulations defines a \(G\)-extension of an original symmetry category. \(G\)-extensions of fusion categories is explored in \cite{Etingof:2009yvg}. It would be interesting to elaborate how the result in \cite{Etingof:2009yvg} can be understood in the SymTFT picture and specify enhanced symmetry categories for given surface operators.
    \item In this paper we give two lattice model examples for codimension-two transitions. In both cases, topological manipulations that specify the Morita equivalences of enhanced categories come from that of the original \(\mathbb{Z}_2\times\mathbb{Z}_2\) symmetry. Mathematically, a Morita equivalence between two enhanced categories \(\mathcal{C}_1\) and \(\mathcal{C}_2\) is specified by a module category over \(\mathcal{C}_1\), and it is not necessarily a module category over the original symmetry. It would be intriguing to find Morita equivalences specified by a topological manipulation defined only for an enhanced symmetry category and construct a lattice model for the corresponding codimension-two transition.
    \item Generalization to fermionic systems or higher dimensions.
\end{itemize}

\section*{Acknowledgements}
The author thanks Kansei Inamura, Ryohei Kobayashi, Kantaro Ohmori, and Yunqin Zheng for useful discussions on this work. 
The author thanks Kansei Inamura, Ryohei Kobayashi, Masatoshi Sato, and Ken Shiozaki for comments on a draft. This work was supported by JST CREST Grant No.~JPMJCR19T2.
T.A.~is supported by JSPS KAKENHI Grant No. 25KJ1557.

\appendix
\section{Field-theory analysis}\label{sec:Z2Z2_QFT}
In this appendix, we give a brief review of a field-theory-based analysis of systems with \(\mathbb{Z}_2\times\mathbb{Z}_2\) symmetries, which we discuss in the main text.

Let \(Z_{\mathcal{X}}[A,B]\) be a partition function of a \(\mathbb{Z}_2^A\times\mathbb{Z}_2^B\) symmetric theory \(\mathcal{X}\) with the two \(\mathbb{Z}_2\) background gauge fields \(A,B\). We define two topological manipulations \(S\) and \(T\) as follows:
\begin{align}
    Z_{S\mathcal{X}}[\hat{A},\hat{B}]&\coloneqq \#\sum_{a,b\in H^1(M_2,\mathbb{Z}_2)}Z_{\mathcal{X}}[a,b](-1)^{\int_{M_2}a\hat{B}+b\hat{A}},\\
    Z_{T\mathcal{X}}[A,B]&\coloneqq Z_{\mathcal{X}}[A,B](-1)^{\int_{M_2}AB},
\end{align}
where \(M_2\) is a spacetime manifold and \(\#\) is a normalized factor depending on the topology of \(M_2\). \(S\) is gauging the \(\mathbb{Z}_2^A\times\mathbb{Z}_2^B\) symmetry, and  \(T\) is stacking a \(\mathbb{Z}_2\times\mathbb{Z}_2\) SPT. In addition to the above two, we have other topological manipulations. The structure of the six gapped phases and topological manipulations between them were studied in \cite{Gaiotto:2020iye, Moradi:2022lqp}.

\begin{figure}[htbp]
    \centering
    \includegraphics[scale=.8]{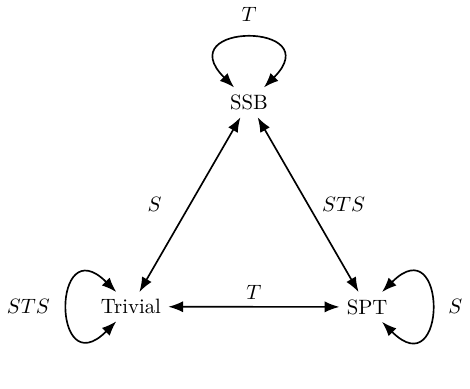}
    \caption{Three gapped phases with \(\mathbb{Z}_2\times\mathbb{Z}_2\) symmetry and the dualities between them. The group structure of the web is \(SL(2,\mathbb{Z}_2)\cong S_3\), see \cite{Li:2023ani}.}
    \label{fig:web_Z2Z2}
\end{figure}
Though \(S\) and \(T\) are both duality operations, the combinations of them are not necessarily duality operations. For example, consider the operation \(ST\), which is defined as
\begin{equation}
    Z_{ST\mathcal{X}}[\hat{A},\hat{B}]=\#\sum_{a,b\in H^1(M_2,\mathbb{Z}_2)}Z_{\mathcal{X}}[a,b](-1)^{\int_{M_2}ab+a\hat{B}+b\hat{A}}.
\end{equation}
This operation describes a \textit{triality} topological manipulation, namely the operation \((ST)^3\) is the identity operation. One can explicitly verify the triality structure by a straightforward calculation.

\section{Else-Nayak procedure}\label{sec:EN}
Once one obtains the expressions of symmetry operators in \((1+1)\)-dimensional systems, one can extract elements of \(H^3(G,U(1))\), which specify the anomaly of the system \cite{Else:2014vma, Kawagoe:2021gqi, Seifnashri:2023dpa}. In this appendix we review the procedure developed by Else and Nayak \cite{Else:2014vma}.

Consider a system with a \(G\) global symmetry whose unitary symmetry operators are \(\{U(g)\}\}_{g\in G}\). We do not assume that \(U(g)\) are on-site operators, but local. Since \(U\) forms a unitary representation, \(U(g)\) obeys the group multiplication law on the closed chain, i.e., \(U(g_1)U(g_2)=U(g_1g_2)\). On the other hand, when we put the system on the open region \(M\) (\(\partial M=\{a,b\}\)), this multiplicity does not necessarily hold. Instead, we have the following relations:
\begin{equation}
    U_M(g_1)U_M(g_2)=\Omega(g_1,g_2)U_{M}(g_1g_2),
\end{equation}
where \(\Omega(g_1,g_2)\) is an operator localized at two edges of the region \(M\). By the associativity, we see that \(\Omega\) must satisfy the following conditions for all \(g_1,g_2,g_3\in G\):
\begin{equation}
    \Omega(g_1,g_2)\Omega(g_1g_2,g_3)=\left[^{U_M(g_1)}\Omega(g_2,g_3)\right]\Omega(g_1,g_2g_3),
\end{equation}
where \(^xy\coloneqq xyx^{-1}\). When one considers the restriction \(\Omega\rightarrow\Omega_{a}\), these relations hold only up to \(U(1)\) phase. Namely, there is a three-cocycle \(\omega\in C^3(G,U(1))\) such that
\begin{align}\label{EN-3-cocycle}
    \begin{split}
        \Omega_a(g_1,g_2)\Omega_a(g_1g_2,g_3)=&\omega(g_1,g_2,g_3)\left[^{U_M(g_1)}\Omega_a(g_2,g_3)\right]\\
        &\times\Omega_a(g_1,g_2g_3),
    \end{split}
\end{align}
where \(\Omega_{a}\) is an operator at the edge \(a\) and comes from the restriction of \(\Omega\). By the associativity for \(\Omega\), one can show that \(\omega\in Z^3(G,U(1))\). Dividing appropriate ambiguities we can define \([\omega(g_1,g_2,g_3)]\in H^3(G,U(1))\), and this is the anomaly three-cocycle of the system.
\paragraph*{Remark.}
Though the expressions of \(\Omega\) and \(\omega\) depend on how one defines the symmetry operator for open regions, the cohomology class \([\omega]\) is independent of the choice of \(\Omega\) and the restriction \(\Omega\to\Omega_{a}\).

\section{Symmetry category enhanced by stacking SPTs}\label{sec:enhance_by_SPT}
In this appendix, we discuss symmetry categories enhanced by stacking SPTs. In \((1+1)\) dimensions, bosonic SPT phases with a finite \(G\) group are classified by the second group cohomology \(H^2(G,U(1))\), which is an Abelian group. We assume that \(H^2(G,U(1))\cong \prod_{l=1}^{M}(\mathbb{Z}/n_{l}\mathbb{Z})\) and denote the generator of the \(l\)-th element of the right hand side by \(\omega_{l}\). Since SPT stacking operations are invertible or realized by FDLUs, the enhanced symmetry category obeys the fusion rule of the group \(G\times H^2(G,U(1))\). To specify the symmetry category, we need the information of \(F\)-symbols, i.e., an 't Hooft anomaly of the group. We show that the anomaly three-cocycle of the enhanced symmetry category is given by 
\begin{equation}\label{eq:omega_SPT}
    e^{2\pi i\sum_{l=1}^{M}\int h^{l}\cup(g^*\omega_{l})},
\end{equation}
where \(h^{l}\) and \(g\) is the background gauge field for \(\mathbb{Z}/n_{l}\mathbb{Z}\) and \(G\), respectively. To show the three-cocycle, we use the procedure developed by \cite{Else:2014vma}, which is reviewed in App.~\ref{sec:EN}. We assume that we have one-dimensional on-site symmetry action \(\{U(g)\}_{g\in G}\), i.e., \(U\) forms a unitary representation of the symmetry group \(G\). We define symmetry operations for the enhanced symmetry as 
\begin{equation}
    \widetilde{U}(g,h=(h^{1},\ldots,h^{M}))\coloneqq U(g)V(h)=U(g)\prod_{l=1}^{M}V_{l}^{h^{l}}.
\end{equation}
Here, we denote the element of \(H^2(G,U(1))\) by \(h=(h^1,\ldots,h^M)\). Consider an open region \(M,\partial M=\{a,b\}\) and fix the representation of \(V(h)\) on \(M\). On the open chain \(V_{l}U(g)V_{l}^{-1}\) is not equal to \(U(g)\), and the difference appears as local operators localized on the edge. We write such a local operator at the edge \(a\) as
\begin{equation}
    \left.V_{l}U(g)V_{l}^{-1}\right|_{a}\eqqcolon \sigma_{a,l}(g)U(g).
\end{equation}
Since \(V_{l}\) describes the stacking SPT operation whose second group cohomology is \(\omega_{l}\), \(\{\sigma_{l}(g)U(g)\}_{g}\) forms a projective representation of \(G\) specified by \([\omega_{l}]\in H^2(G,U(1))\). Note that \(\{\sigma_{a,l}(g)\}_{g}\) also forms a representation of \(G\), which is projective in general. However, we can redefine it so that it forms a genuine representation of \(G\), and we take the genuine representation in the following. By using \(\sigma_{a,l}\), we write the local operator \(\Omega_{a}\), which appears in \eqref{EN-3-cocycle} as 
\begin{align}
    \begin{split}
        \Omega_{a}\left((g_1,h_1),(g_2,h_2)\right)=&U(g_1)V_{M}(h_1)U(g_2)V_{M}(h_2)\\
        &\times \left(U(g_1g_2)V(h_1h_2)\right)^{-1}\left.\right|_{M\rightarrow a}\\
        =&U(g_1)\left(\prod_{l=1}^{M}\sigma_{a,l}(g_2)^{h_1^l}\right)U(g_1)^{-1}\\
        =&e^{2\pi i \sum_{l=1}^{M}h_1^l\omega_{l}(g_1,g_2)}\prod_{l=1}^{M}\sigma_{a,l}(g_2)^{h_1^l}.
    \end{split}
\end{align}
Then from \eqref{EN-3-cocycle}, we obtain the following anomaly three-cocycle:
\begin{equation}
    \omega\left((g_1,h_1),(g_2,h_3),(g_3,h_3)\right)=e^{2\pi i\sum_{l=1}^{M}h_1^l\omega_{l}(g_2,g_3)}.
\end{equation}
This completes the proof of \eqref{eq:omega_SPT}.

\section{Compact boson calculation}\label{sec:cptbsn}
In this appendix, we provide a detailed analysis of the IR behavior of the lattice model \eqref{Z2Z2_ham}. Again, we consider the three self-duality lines in Fig.~\ref{fig:Z2Z2_phase}. First, let us consider the point \(J_0=J_S, J_1=0\), which is described by two decoupled Ising CFTs. We have the \(U(1)_4\) (free Dirac) CFT after implementing the manipulation \(ST\) \cite{Hsin:2020nts}. More explicitly, we have
\begin{align}
    \begin{split}
        Z_{ST(\text{Ising}^2)}[\hat{A},\hat{B}]
        &\coloneqq \#\sum_{a,b}Z_{\text{Ising}^2}[a,b](-1)^{\int_{M_2}ab+a\hat{B}+b\hat{A}}\\
        &=Z_{U(1)_4}[\hat{A},\hat{B}].
    \end{split}
\end{align}
However, the calculation in Sec.~\ref{sec:Z2Z2_model} does not specify how the dual \(\mathbb{Z}_2^{\hat{A}}\times\mathbb{Z}_2^{\hat{B}}\) symmetry couple to the \(U(1)_4\) CFT. By an explicit computation, we can see that these two \(\mathbb{Z}_2\) symmetries are described by combinations of two \(\mathbb{Z}_2\) symmetry, the \(\mathbb{Z}_2^S\) shift symmetry and the \(\mathbb{Z}_2^C\) charge conjugation symmetry of the \(U(1)_4\) CFT. We denote the \(\mathbb{Z}_2\) gauge fields for each symmetry by \(A^S,A^C\). Specifically, when we compute the torus partition function with defects of the \(ST(\text{Ising}^2)\), we obtain 
\begin{gather}
    Z_{ST(\text{Ising}^2)}[\hat{A},0]=Z_{U(1)_4}[\hat{A}^C],\\
    Z_{ST(\text{Ising}^2)}[0,\hat{B}]=Z_{U(1)_4}[\hat{B}^C],\\
    Z_{ST(\text{Ising}^2)}[\hat{A},\hat{A}]=Z_{U(1)_4}[\hat{A}^S],
\end{gather}
and other twisted partition functions with off-diagonal defect insertions vanish. 
On the other hand, the non-invertible symmetry in the \(U(1)_4\) CFT is mapped to the winding \(\mathbb{Z}_2\) symmetry in the conpact boson.
According to the lattice analysis in Sec.~\ref{sec:Z2Z2_enhance}, two dual \(\mathbb{Z}_2\) symmetries in \(ST(\text{Ising}^2)\) and the winding \(\mathbb{Z}_2\) symmetry carry the type III anomaly and no other mixed anomalies.
From them, we see that 
\((\mathbb{Z}_2^{\hat{A}},\mathbb{Z}_2^{\hat{B}})=((\mathbb{Z}_4^S\times\mathbb{Z}_2^C)_{\text{diag.}},(\mathbb{Z}_2^C\times\mathbb{Z}_4^S)_{\text{diag.}}),~\text{or}~((\mathbb{Z}_2^C\times\mathbb{Z}_4^S)_{\text{diag.}},(\mathbb{Z}_4^S\times\mathbb{Z}_2^C)_{\text{diag.}})\), where \((\mathbb{Z}_4^S\times\mathbb{Z}_2^C)_{\text{diag.}}\) is for example generated by the generator of the \(\mathbb{Z}_2^C\) symmetry and the generator of the \(\mathbb{Z}_4^S\) symmetry.
Note that we cannot distinguish two \(\mathbb{Z}_2\) defects \(\mathbb{Z}_2^C\) and \((\mathbb{Z}_4^S\times\mathbb{Z}_2^C)_{\text{diag.}}\) at the level of defect partition function on the torus.

As we change the value of the interaction \(J_1\), the radius of the compact boson and the orbifolded boson are continuously deformed. Thus, the two global \(\mathbb{Z}_2\) symmetries \eqref{Z2Z2_sym} of the (\(ST\) operated) Hamiltonian \(H[J_0=J_1,J_S]\) match \(\mathbb{Z}_2^C\) and \((\mathbb{Z}_2^S\times\mathbb{Z}_2^C)_{\text{diag.}}\) in the IR. The fact that the diagonal \(\mathbb{Z}_2\) subgroup of the two is the \(\mathbb{Z}_2^S\) shift symmetry is consistent with the analysis in Sec.~\ref{sec:Z2Z2_model}.

Let us see the self-triality structure of the KT CFT. On the torus, we see that 
\begin{align*}
        Z_{ST(\text{KT})}
    =&\frac{1}{4}\sum_{a,b}Z_{\text{KT}}[a,b](-1)^{\int ab}\\
    =&\frac{1}{4}\left(\sum_{a=b}Z_{\text{KT}}[a,b]+\sum_{a\neq 0}Z_{\text{KT}}[a,0]\right.\\
    &+\left.\sum_{b\neq 0}Z_{\text{KT}}[0,b]+\sum_{\substack{a\neq 0,b\neq 0,\\a\neq b}}Z_{\text{KT}}[a,b](-1)^{\int ab}\right)\\
    =&\frac{1}{4}\left(2Z_{\text{KT}/\mathbb{Z}_2^S}\right.\\
    &\left.+2\cdot 2\left(\left|\frac{\eta(\tau)}{\theta_2(\tau)}\right|+\left|\frac{\eta(\tau)}{\theta_4(\tau)}\right|+\left|\frac{\eta(\tau)}{\theta_3(\tau)}\right|\right)+0\right)\\
    =&\frac{1}{2}Z_{SU(2)_1}+\left|\frac{\eta(\tau)}{\theta_2(\tau)}\right|+\left|\frac{\eta(\tau)}{\theta_4(\tau)}\right|+\left|\frac{\eta(\tau)}{\theta_3(\tau)}\right|\\
    =&Z_{SU(2)_1/\mathbb{Z}_2^C}\\
    =&Z_{\text{KT}}.
\end{align*}
Here, \(\theta_i(\tau)\coloneqq \theta_i(\tau,z=0)\) are the Jacobi theta functions and \(\eta(\tau)\) is the Dedekind eta function. From the second line to the third line, we used the fact that \(\text{KT}/\mathbb{Z}_2^S\) and the partition function with off-diagonal defects vanish as in \(U(1)_4\) case. In the last line, we used the fact that the \(\mathbb{Z}_2^C\) orbifolded theory of the \(SU(2)_1\) CFT is the KT CFT. The self-triality of the KT CFT was studied in \cite{Thorngren:2021yso}, and later further explored in \cite{Lu:2022ver}.

\section{Detail discussion on \texorpdfstring{\(\mathbb{Z}_n\times\mathbb{Z}_n\)}{Zn x Zn} symmetry}\label{sec:ZnZn_QFT}
In this appendix, we give the proof of the Claim \ref{claim} in Sec.~\ref{sec:SPT_transition} by explicitly constructing the topological manipulation. In \((1+1)d\), we have \(p\) distinct bosonic SPT phases with a \(\mathbb{Z}_p\times\mathbb{Z}_p\) symmetry, which are labeled by integer \(k~\text{mod}~p\) with the partition function 
\begin{equation}
    Z_{\text{SPT}_k}[A,B]=e^{\frac{2\pi i}{p}\int_{M_2} kAB},
\end{equation}
where \(M_2\) is a spacetime manifold. We call \(\text{SPT}_k\) a level-\(k\) SPT. We define two topological manipulations \(S\) and \(T_1\) for a theory \(\mathcal{X}\) as
\begin{gather}
    Z_{S\mathcal{X}}[A,B]\coloneqq \sum_{a,b\in H^1(M_2,\mathbb{Z}_p)}e^{\frac{2\pi i}{p}\int_{M_2}aB+bA}\\
    Z_{T_1\mathcal{X}}[A,B]\coloneqq Z_{\mathcal{X}}[A,B]\,e^{\frac{2\pi i}{p}\int_{M_2}AB}.
\end{gather}
\(S\) is gauging the \(\mathbb{Z}_p\times\mathbb{Z}_p\) symmetry and \(T_1\) is stacking the level-one SPT phase.\footnote{We suppress the overall normalization throughout this appendix.} First, we show the following proposition. 
\begin{prop}\label{prop_Zpdual}
    Let \(p\) be an odd prime integer. There is a duality operation between any two \(\mathbb{Z}_p\times\mathbb{Z}_p\) SPTs with different levels. 
\end{prop}
To show the proposition, it is enough to construct the duality between \(\text{SPT}_0\) and \(\text{SPT}_1\). One can see that a duality operation \(f=S\circ T_{1}^{-1}\circ S\circ T_{1} \circ S\) realizes such operation.
\bigskip

How does the manipulation \(f\) act on other SPT phases? From the definition of \(S\), we see that
\begin{align}
    \begin{split}
        Z_{S(\text{SPT}_{k})}[A,B]&=\sum_{a,b}e^{\frac{2\pi i}{p}\int kab+aB+bA}\\
        &=\sum_{a}\delta(-ka+A)e^{\frac{2\pi i}{p}\int aB}\\
        &=e^{\frac{2\pi i}{p}\int \frac{AB}{k}}=Z_{\text{SPT}_{1/k}}[A,B].
    \end{split}
\end{align}
Here, we used that \(p\) is prime. Then we find the topological manipulation \(S\) change the level of SPTs from \(k\) to \(1/k\). Under the manipulation \(f\), level-\(k\) (\(k\neq 0,1\)) SPT phase is mapped as
\begin{equation}\label{SPTk_f}
    k\mapsto \frac{1}{k}\mapsto \frac{1}{k}-1 \mapsto \dfrac{1}{\dfrac{1}{k}-1}\mapsto \dfrac{1}{\dfrac{1}{k}-1}+1 \mapsto \left(\dfrac{1}{\dfrac{1}{k}-1}+1\right)^{-1}.
\end{equation}
Then the self-duality condition under \(f\) is
\begin{equation}
    \frac{1}{q-1}+1=q \quad \text{mod}~p,
\end{equation}
where \(q=1/k\). The condition is equivalent to \(q(q-2)\equiv 0\). Therefore, an SPT with level-\(1/2\) is self-dual under the manipulation \(f\).

In the proof of Proposition \ref{prop_Zpdual}, the point is that we can define SPTs with level \(1/k\). Such a dividing is not defined for general integers \(n\). Nevertheless, we can use the same duality operation for some SPT pairs. By simple consideration, we find that this happens when the difference of the SPT level is coprime to \(n\). Therefore, we obtain the following statement.
\begin{cor}
    Consider two SPT phases with a \(\mathbb{Z}_n\times\mathbb{Z}_n\) symmetry of levels \(k\) and \(k^\prime\). There is a duality operation between these two SPT phases if \(k-k^\prime\) is coprime to \(n\).
\end{cor}
This is the Claim \ref{claim} in Sec.~\ref{sec:SPT_transition}. We can apply these operations to concrete lattice models \eqref{eq:ZnZn_SPT} and \eqref{eq:ZnZn_SSB}. Now we see that self-dual critical points between two \(\mathbb{Z}_n\times\mathbb{Z}_n\) SPT phases are obtained from the two decoupled \(\mathbb{Z}_n\) clock models by gauging with discrete torsion twists. This gives the generalization and interpretation of the non-local transformations discussed in \cite{Tsui:2017ryj}. 

\bibliography{ref}
\end{document}